\documentclass[12pt, draftclsnofoot, onecolumn]{IEEEtran}
\usepackage{amsmath,amsfonts}
\usepackage{algorithmic}
\usepackage{algorithm}
\usepackage{array}
\usepackage[caption=false,font=normalsize,labelfont=sf,textfont=sf]{subfig}
\usepackage{textcomp}
\usepackage{stfloats}
\usepackage{url}
\usepackage{verbatim}
\usepackage{graphicx}
\usepackage{cite}
\usepackage{gensymb}
\usepackage{hyperref}
\begin{document}

\title{Towards the THz Networks in the 6G Era}

\author{
	\IEEEauthorblockN{
		Qian Ding\IEEEauthorrefmark{2}, 
		Jie Yang\IEEEauthorrefmark{3},
		Yang Luo\IEEEauthorrefmark{2},
		and Chunbo Luo\IEEEauthorrefmark{2}\IEEEauthorrefmark{1}}\\ 
	\IEEEauthorblockA{\IEEEauthorrefmark{2} University of Electronic Science and Technology of China, Chengdu, China}\\
	\IEEEauthorblockA{\IEEEauthorrefmark{3} Huawei Technologies Co., Ltd, Xi'an, China\\Email: c.luo@uestc.edu.cn (Chunbo Luo)}
} 


\maketitle

%

\section{Introduction}
As the fifth generation (5G) wireless communication networks being rapidly deployed and commercialized since 2020, the communication communities turn their gaze to the future sixth generation (6G) networks expected to materialize in the 2030s. In comparison to its predecessor, 6G is anticipated to achieve superior performance across various metrics, including spectral efficiency, energy efficiency, connectivity, coverage, latency and security. Amidst the speculative studies surrounding 6G, a consensus is emerging: 6G is projected to be inherently human-centric\cite{dang2020should,you2021towards}. This distinctive trait of 6G introduces new requirements for user experience and thereby bring forward a host of appealing applications such as extended reality (XR), telemedicine and digital twins, demanding exceptionally high bandwidth and strong perception and communication capabilities.

\begin{figure*}[t]
	\centering
	\includegraphics[width=\columnwidth]{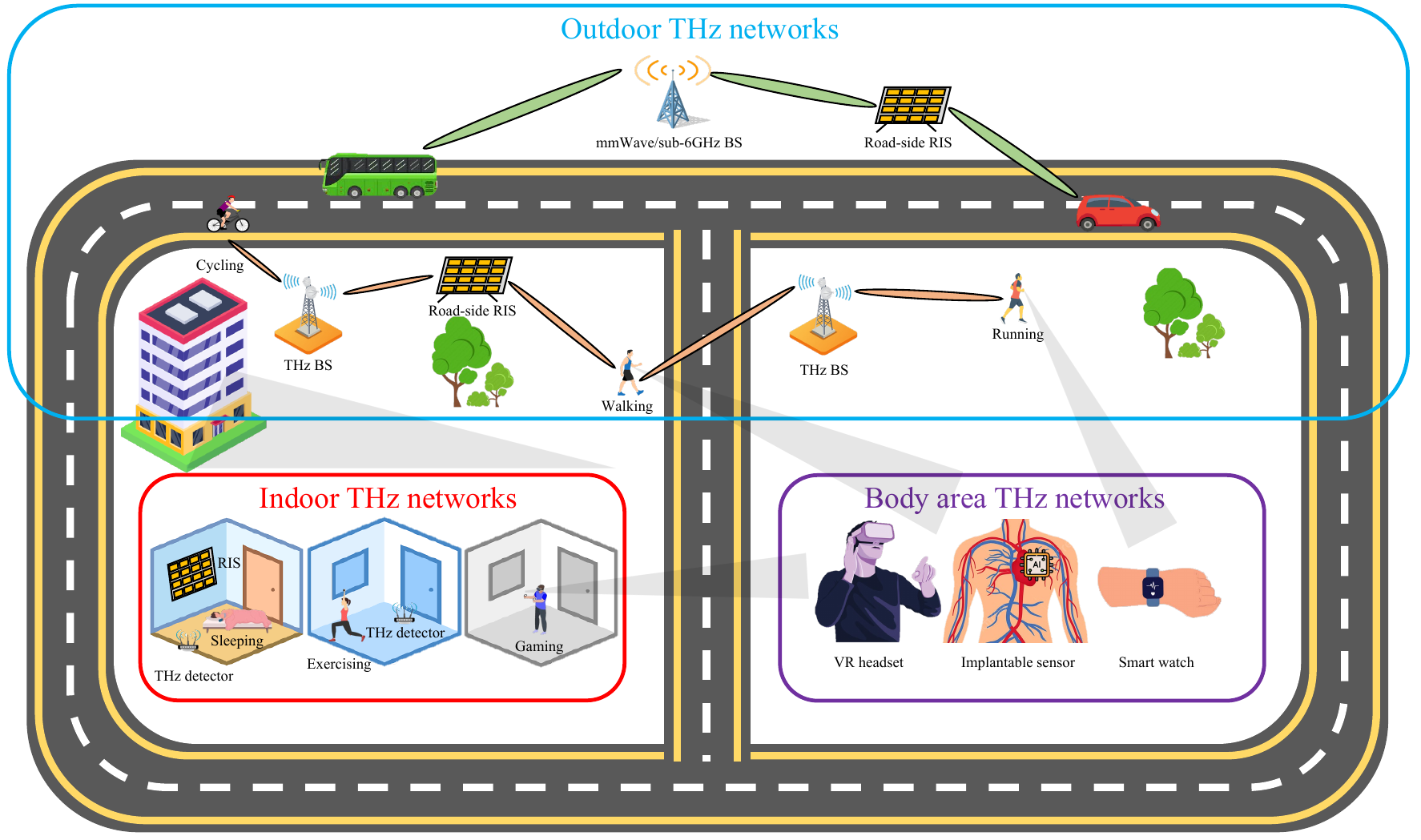}
	\caption{Illustration of outdoor, indoor and body area THz networks}
	\label{fig:Overall display of three types of networks}
\end{figure*}

Terahertz (THz) technology, spanning from 0.1THz to 10THz and abundant in unexplored frequency bands, emerges as a pivotal enabler for the transition from 5G to 6G\cite{chaccour2022seven}. Beyond providing high-speed data transmission rates, THz band, thanks to its unique quasi-optical nature, offers strong capabilities in high-precision sensing, imaging and localization. Thus, the convergence of communications, sensing, computing, control and localization is likely to be achieved with THz, which redifines 6G wireless communication networks and transforms them into versatile platforms rather than mere data transmission. On the other hand, benefiting from the non-ionizing radiation properties, THz holds promise for non-invasive and non-contact inspections, especially useful for health examinations and continuous sensing. Given these attributes, THz technologies emerge as pivotal tools for outdoor, indoor and human body area sensing and interaction. This commentary thus focuses particularly on the role of THz in three typical scenarios in the 6G era: outdoor networks, indoor networks and body area networks, as displayed in Fig. \ref{fig:Overall display of three types of networks}.
\subsection{Outdoor networks}
We, humans, are the core participants in the outdoor activities and acquiring detailed information about human behaviors is pivotal for safety and planning purpose. Compared to the swift movement of autonomous vehicles sensed by the millimeter (mmWave) band or sub-6GHz, human motion exhibits a relatively slower pace, making it conducive for tracking by THz base stations (BSs). The high-precision sensing ability of THz enables more sensitive and accurate detection of human macro-mobility such as walking, running and cycling as well as micro movements such as user orientation, than the fully-fledged global positioning system (GPS)\cite{chaccour2022seven}. However, the practical implementation of THz for outdoor environment involves several challenges. Firstly, the propagation distance of THz waves is inherently limited due to their high frequency and susceptibility to molecular absorption effects. Consequently, beamforming technology is extensively adopted to overcome the restricted range through focusing the transmit power towards the desired user directions. Given the ultra-narrow THz beam, the design of advanced beam tracking algorithms becomes pivotal, considering the continuous movement of human participants on the roads. Secondly, the development of highly precise prediction models is imperative to characterize diverse human behaviors accurately. Thirdly, in densely populated regions where sub-6GHz and mmWave frequency bands dominate, ensuring compatibility and cooperation between THz networks and these lower frequency bands becomes crucial. Last but not least, from a communication perspective, THz band can support ultra-high data rates. In this sense, configuring THz systems in an integrated sensing and communication (ISAC) framework appears favorable. Nonetheless, this integration poses challenges in network resource allocation between the two functionalities, demanding careful management to avoid compromising the user experience.
\subsection{Indoor networks}
From the propagation perspective, THz waves are more suitable for the indoor environment due to shorter transmission distance and lower water vapor absorption. The introduction of THz indoor networks opens up a wide spectrum of opportunities. Notably, THz technology can achieve centimeter level localization precision or $\degree$ level resolution given that the electromagnetic wave are well processed. Leveraging this high resolution sensing, accurate indoor positioning that cannot be achieved by conventional GPS becomes possible. Moreover, THz waves hold promise for assessing indoor activities such as exercising, working, sleeping, and even monitoring vital signs. These applications lay the foundation for various future scenarios such as smart homes and telemedicine. Nonetheless, several critical issues necessitate urgent attention. Firstly, security capacity ought to be strengthened for the health-related data, without bearing the data processing efficiency. Additionally, the indoor scenes are enclosed and consequently susceptible to the human activity, and as a result the line-of-sight (LoS) paths are generally difficult to maintain. Innovative technologies like reconfigurable intelligent surfaces (RISs) are being explored to enhance THz performance and enable holographic data transmission in such scenarios\cite{chen2022electrically}. Furthermore, human behavior in indoor environments significantly differs from that of pedestrians on roads. This contrast presents novel challenges to design dedicated prediction models and sensing algorithms tailored for indoor dynamics.
\subsection{Body area networks}
The non-ionizing nature of THz radiation, coupled with its remarkable ability to penetrate non-polarized objects, positions THz as an appealing technology for on-body devices like wearable devices, virtual reality (VR) headsets, and implantable sensors. Thanks to the development of advanced materials such as graphene and device miniaturization, wireless body area networks (WBANs) comprised of those on-body devices gradually become a research hotpot\cite{suzuki2016flexible}. With the help of WBAN, a comprehensive detection of human body is likely to be achieved to enable applications ranging from real-time health monitoring in biomedical fields to immersive experiences like XR. Meanwhile, the sensitivity and confidentiality of physical health data necessitate extensive exploration into secure communication and privacy protection schemes like federated learning. Additionally, the battery life of on-body devices, particularly implantable sensors, demands careful consideration, as inadequate longevity could significantly diminish the practicality and sustainability of WBANs. Addressing this energy-limiting aspect calls for widespread research into devices with ultra-low power consumption. Furthermore, ensuring wearing comfort is paramount for enhancing the quality of user experience. Therefore, the factors such as stretchability, flexibility and reliability are equally crucial for the future WBAN devices.
\subsection{Synergy with other technologies}
RIS and artificial intelligence (AI) are widely regarded as the other two key enabling technologies for 6G\cite{dang2020should}, and THz technology exhibits a compelling synergy with these advancements. On one hand, THz waves, characterized by their short wavelength and pencil beam, often encounter challenges in establishing LoS links and consequently require additional links to ensure stable THz connectivity. The advent of RIS, a revolutionary architecture comprised of numerous meta elements, presents a solution by creating supplementary paths and reconfiguring the communication environments via adjusting the magnitude and phase of reflecting signals. By densely and properly deploying RISs in the THz networks, ubiquitous coverage and reliable links can be achieved. On the other hand, the intermittent nature and uncertainty of THz channel impose additional challenges to the initial access and network reliability. Traditional channel estimation methods designed for mmWave bands often fall short when applied to THz frequencies. Consequently, researchers have turned their attention to leveraging the strong learning capabilities of AI methods. Recently, generative learning, due to its ability to reason, learn, adapt, and act independently, has emerged as a promising approach for capturing the distinctive characteristics of THz channels and predicting channel state information (CSI)\cite{you2021towards}.
\section{Conclusion}
This commentary dedicates to envision what role THz is going to play in the coming human-centric 6G era. Three distinct THz network types including outdoor, indoor, and body area networks are discussed, with an emphasis on their capabilities in human body detection. Synthesizing these networks will unlock a bunch of fascinating applications across industrial, biomedical and entertainment fields, significantly enhancing the quality of human life. 
\section*{declaration of interests}
The authors declare no competing interests.
%
\bibliographystyle{IEEEtran}
\bibliography{ref}

\vfill

\end{document}